\documentclass[11pt]{article}
\pdfoutput=1
 \usepackage{textcomp}  
\usepackage{amsmath,amssymb} 
\usepackage{epsfig,latexsym}
    \usepackage{graphicx}
\newcommand{\sect}[1]{\setcounter{equation}{0}\section{#1}}

\def\al {\alpha}

\def\ba{\begin{eqnarray}}
 \def\bbb{background~}
\def\bbbb{backgrounds~}
\def\be{\begin{equation}}
\def\bi{\bibitem}
\def\bm{\boldmath}
\def\bxi{\bm{ $\xi$}\ubm}
\def\bt {\beta}

\def\Bar {\overline}
\def\BB{{\cal B}} 
\def\BD{\Bar D}
\def\Bg{\Bar g} 

\def\BGa{\Bar\Ga}
\def\BR{\Bar R}

\def\cd{\cdot}

\def\coo{coordinates~}

\def\de{\delta}

\def\di{\partial}
\def\tLL{\dot{\LL}}

\def\DD{{\cal D}}
\def\De{\Delta}

\def\ea{\end{eqnarray}} 
\def\ee{\end{equation}}
\def\eee{equation~}
\def\eeee{equations~}

\def\ep{\epsilon} 
\def\eq{\equiv~}

\def\fr{\frac}

 \def\ga{\gamma}

\def\Ga{\Gamma}

\def\GR{General Relativity~}
\def\ha{\frac{1}{2}~}
 \def\hg{\hat g}

\def\in{\infty}

\def\ka{\kappa}

\def\la {\lambda}
\def\lb{\label} 
\def\lhs{left-hand side~}

\def\lll{\left(}

\def\La {\Lambda}
\def\LL{{\cal L}}
\def\LLL{\left[}

\def\mn{\mu\nu}

\def\na{\nabla}

\def\nn{\nonumber}  

\def\nnn{\noindent}

\def\np{\newpage}

\def\OO {{\cal O}}

\def\po{\pounds}

\def\rhs{right-hand side~}
\def\rh{\rho}

\def\rrr{\right)}
\def\rs{\rho\si}

\def\Ra{\Rightarrow}
\def\RRR {\right]}

\def\si{\sigma}
\def\sq{\sqrt}
\def\sqg{\sqrt{-g}}

\def\sss{spacetime~}
\def\ssss{spacetimes~}

\def\sup{superpotential~}
\def\supp{superpotentials~}

\def\Sc{Schwarzschild~}

\def\td{\tilde}

\def\tha{\ts{\ha}}

\def\tLL{\tilde\LL}

\def\ts{\textstyle}

\def\tW{\td W}

\def\ubm{\unboldmath}

\def\us{\underset}

\def\vna{\vec{\na}}
\def\vs{\vskip 0.5 cm}

\def\we{\wedge}
\def\wrt{with respect to~}

\def\1k{\fr{1}{\ka}}

\def\2k{\fr{1}{2\ka}}

\title{{\bf  Superpotentials from variational derivatives rather than\\Lagrangians  in relativistic theories of gravity }}
\author{ Joseph Katz$^1$\thanks{email: jkatz@phys.huji.ac.il} \,\,and Gideon I.  Livshits$^2$\thanks{email:
livshits.gideon@mail.huji.ac.il}\\
 \\  {\it$^1$The Racah Institute of Physics and $^2$Institute of Chemistry}
\\
 \\{\it The Hebrew University, Givat Ram, 91904 Jerusalem, Israel}}
 
\begin{document}
\maketitle

\begin{abstract} 
\setlength{\baselineskip}{20pt plus2pt}
The prescription of Silva to derive superpotential equations from variational derivatives rather than from Lagrangian densities is applied to theories of gravity derived from Lovelock Lagrangians in the Palatini representation. Spacetimes   are without torsion and isolated  sources of gravity are minimally coupled.  On a closed boundary of spacetime, the metric is given and the connection coefficients are those of Christoffel. We derive    equations for the superpotentials in these conditions. The \eeee are easily integrated and we give the general expression for   all superpotentials associated with Lovelock Lagrangians. We find, in particular, that in Einstein's theory, in  any number of dimensions, the superpotential, valid at spatial and at null infinity, is that of Katz, Bi\v c\'ak and Lynden-Bell, the KBL superpotential. We also   give explicitly the superpotential for Gauss-Bonnet theories of gravity. Finally, we find a simple expression for the \sup of Einstein-Gauss-Bonnet theories with an anti-de Sitter background: it is {\it minus} the KBL superpotential, confirming, as it should,  the calculation of the total mass-energy of spacetime at spatial infinity  by Deser and  Tekin.  

\vs

\vs\vs\vs
PACS number(s): 04.20.-q , 04.20.Cv, 04.20.Fy , 04.50.-h

\end{abstract} 
\np

\setlength{\baselineskip}{20pt plus2pt}
\sect{Introduction}
\vs
{\it (i) A view on \supp}
\vs
In electromagnetism, one of Maxwell's \eeee relates the stationary electric field $\vec E$ to the density of charges $\rho_e$: $\vna\cd\vec E=4\pi \rho_e$. From this follows that the total charge $Q$ responsible for the field is equal to the flux of  $\vec E$ through a closed surface surrounding the sources. $\vec E$ is the electric force acting on a unit test charge $q$.   In Newton's theory of gravitation the gravitational field $\vec G$ is related to the density of matter $\rho_m$ in a similar way: $\vna\cd\vec G=  - 4\pi \rho_m$. Thus the flux of $ - \vec G$ through a closed surface surrounding the source responsible for  $\vec G$  is equal to the total mass $M$. $\vec G$ is the gravitational force acting on a unit test mass $m$. In Einstein's theory of gravitation things get slightly more complicated. First there is a change of meaning: total mass is   now total mass-energy $Mc^2$. Second,  given a localized source of gravity, part of the total mass-energy is in the gravitational field itself though its density is not defined nor is it possible in general\footnote{See however \cite{KLB1} and \cite{KLB2}.} to even disentangle the total gravitational field energy from the total mass-energy of spacetime. Nevertheless, an isolated amount of matter together with its gravitational field appears from a great distance as a point-like source of gravitation, possibly spinning, and like in Newton's theory and in electromagnetism its mass-energy is also equal to a flux across a closed surface at spatial infinity. However a surface element in spacetime is a 2-index anti-symmetric  tensor so that instead of  a vector ($\vec E$ or $\vec G$)  we have a two index antisymmetric tensor   whose flux across the surface at infinity equals   mass-energy. The tensor   is commonly called  the superpotential. It is worth noting that the flux includes the   energy of the sources \cite{KBL}.

Unfortunately the superpotential is not as well defined as in classical field theory. Given a Lagrangian from which Einstein's equations are derived, there exists a unique superpotential associated with that Lagrangian\footnote{This is not shown here. It can be derived from the Belinfante-Rosenfeld identities, see \cite{KBL}, without invoking boundary conditions or    field equations.}. The trouble is that the Lagrangian density itself is only defined up to a divergence and is thus not unique.  That is the reason why efforts have been made to obtain the superpotential not from the Lagrangian but from the variational derivatives of the Lagrangian\footnote{And not from field equations as is sometimes stated, see for instance \cite{FFR}.}. These are insensitive to additional divergences.
\vs
{\it (ii) Superpotentials derived from variational derivatives}
\vs
 Silva \cite{Si} suggested a method to obtain     superpotentials from variational derivatives. Previous works, in particular those of Joe Rosen \cite{Ro} and  of  Anderson and Torre \cite{AT} aimed at   obtaining conserved quantities from the field equations. Here the emphasis is on superpotentials and there is no need for field equations. The fluxes we are after are in addition  not necessarily conserved like, for instance, the Bondi mass \cite{Bo} at null infinity. In the last decades relativists   have become   interested in spacetimes with more than $4$ dimensions or which far from the sources are not flat; they  become, in particular, anti-de Sitter\footnote{The mass-energy in such \ssss have been calculated in a series of fine works by Deser and collaborators \cite{AD}, \cite{DT1}, \cite{DT2} to which we shall come back later. See also the paper by Petrov \cite{Pe}.}. The proper way to deal with this situation as well as  in keeping the ability to calculate, for instance mass-energy, in asymptotically flat \ssss  {\it in spherical coordinates}, is  by introducing a background metric\footnote{Elsewhere backgrounds  appears in the guise of ``counter-terms" like in \cite{HIM}  or  a ``regularization" procedure  as in  \cite{KO}. A detailed discussion on the \bbb formalism is given in \cite{KBL}. See also an interesting discussion in \cite{JS2}.} \cite{JS1}.   Ferraris, Francaviglia and Raiteri \cite{FFR}, who   generalized, to some extent,   Silva's idea  were concerned with the  variations of conserved quantities.  
 \vs
{\it (iii) What is done in this paper}
\vs
 Silva's Lagrangian approach is summarized in Section 2. Julia and Silva \cite{JS2} applied the method to Einstein's relativistic theory of gravitation and found that   given the asymptotic  metric components at spatial or at null infinity, the unique answer is the KBL superpotential \cite{KBL}. Silva's prescription applies to variational derivatives  that depend at most on first order derivatives of the field components and whose   Lie derivatives   with respect to some arbitrary vector field depend also at most on first order derivatives of this vector field. These conditions are met in a $GL(D,{\bf R})$ formulation of Einstein's theory. Ferraris, Francaviglia and Raiteri \cite{FFR} extended Silva's prescription to Lagrangians that contain higher order derivatives of the fields but, in the Lie derivatives, do not contain   higher orders of derivatives of the vector field. Their work renders Silva's prescription applicable to Einstein's ordinary equations in terms of the metric components and the \sup we derived that way is again the KBL one. In Section 3 we extend Silva's prescription to the case in which Lie derivatives contain second order derivations of the vector field. This makes the method applicable to variational derivatives of Lovelock Lagrangians\footnote{The method is equally applicable to    more general higher order derivative Lagrangians.}  \cite{Lo} (Section 4) in the Palatini representation in which   the    metric and   the symmetric connection coefficients are   the independent fields.  We find the equation of the \sup and its {\it solution}.  
The superpotential of Einstein's gravity theory in $N$ dimensions is, as expected, the KBL superpotential (Section 5). The \sup for Gauss-Bonnet variational derivatives, derived in Section 6, is far more complicated that the KBL superpotential. The \sup of Einstein-Gauss-Bonnet theories is a linear combination of both. In particular, the \sup with an anti-de Sitter background is {\it proportional} to the KBL superpotential. The one associated with the spherically symmetric solution given in \cite{AD} is {\it minus} the KBL superpotential  and on a sphere at spatial infinity this coincides with the \sup found by Deser and  Tekin \cite{DT2}.     
 \vs
 {\it (iv) A   word about identities }
\vs
It may be noticed that nowhere are field equations used. From the beginning to  the end we play with identities. That is not altogether surprising. Globally conserved or non-conserved quantities are constants or functions that appear in the  boundary conditions.  
 
  It is also interesting to note that the derivation of the \sup from a Lagrangian via Noether identities, like it is done in \cite{Ka} or \cite{KBL}, is very different from the  derivation from variational derivatives.  We have not yet tried  to relate the two methods; the relation appears, to us at least, somewhat mysterious.  
 \vs
 {\it (v) What has this to do with conservation laws? }
\vs
This paper is  far removed from conservation law considerations and is more of a mathematical character.   It is therefore perhaps useful to reconnect  superpotentials to conservation laws before we move on. The superpotential is   an antisymmetric tensor density, linear in some arbitrary displacement vector say\footnote{For a definition of notations see Section 2 below.}\bxi\,, defined on the boundary of the domain, usually at spatial or null infinity where spacetime   identifies with the background that does not have to be flat. The relation to ``conservation laws" is as follows, at least in Einstein's theory of gravitation. The ordinary divergence of that antisymmetric tensor density is a divergenceless vector density, say $J^\rh$. If one uses Einstein's field  equations, and this is one of the rare occasions where they are used in this paper, it was found\footnote{When \sss is flat far from the sources.} \cite{KBL} that $J^\rh= \sqg T^\rh_\si \xi^\si$ plus    terms    which in the linear approximation are negligible. If, in particular,  the \bbb has a Killing field of time translations like in a flat spacetime, then one clearly sees   that the   total  flux of  the  conserved current is related to mass-energy conservation. However, if the background possesses no Killing field then $J^\rh$ is a conserved vector with no    obvious physical meaning. So much for conservation laws and the relevance of superpotentials.
 \sect{The Silva prescription}

 {\it (i) Elements}
 \vs
  In what follows we are given a set of tensors on a curved spacetime with components $y_A(x^\la)$ and a Lagrangian density\footnote{In Einstein's theory of gravity which is in $4$ dimensions, Greek indices go from $0$ to $3$. Latin indices from $1$ to $3$. The signature of the metric $g_{\mn}$ is $-2$ and $g$ is its determinant. Covariant derivatives are indicated by a $D$; partial derivatives by a $\di$. The permutation symbol in 4 dimensions is $\ep_{\mu\nu\rho\si}$ with $\ep_{0123}=1$ and in 3 dimensions by $\ep_{klm}$ with $\ep_{123}=1$.  The $4$-volume element $dV_\mu=\fr{1}{3!}\ep_{\mu\nu\rho\si}dx^\nu\we dx^\rho\we dx^\si$, a surface element in $4$ dimensions $dS_{\mn}=\fr{1}{2}\ep_{\mu\nu\rho\si} dx^\rho\we dx^\si$ and in $3$ dimensions $dS_k=\fr{1}{2}\ep_{0klm}dx^l \we dx^m=\fr{1}{2}\ep_{klm}dx^l\we dx^m$. The Sections on Einstein's gravity and Gauss-Bonnet gravity is in $N$-dimensions to which these definitions extend in an obvious way. In those Sections, Greek indices run from $0$ to $N-1$ and the signature of the metric is $-(N-2)$.}  
 \be
 \LL= \LL(y_A, D_\si y_A).
 \lb{21}
 \ee
 The particularity of $\LL$ is that   variational derivatives,  
 \be
 \LL^A\eq\fr{\di\LL}{\di y_A} - D_\si\fr{\di\LL}{\di (D_\si y_A)}, 
 \lb{22}
 \ee
  contain only  first order derivatives of $y_A$.  Notice that partial derivatives \wrt $y_A$ are at $D_\si y_A$ constant and vice versa.  
 The variation of $\LL$ \wrt $y_A$ can thus  be written in this form:
 \be
 \de\LL=\LL^A \de y_A+ D_\si\LLL\fr{\di\LL}{\di(D_\si y_A)} \de y_A\RRR=\LL^A \de y_A+\di_\si\LLL\fr{\di\LL}{\di(D_\si y_A)} \de y_A\RRR. 
  \lb{23}
 \ee
 The spacetime has here the passive role of a  \bbb so that  $\de D=D\de$. The Lie derivative of a Lagrangian density\footnote{One should pay attention to the fact that we deal with vector densities and that ordinary divergences of such vectors as well as of  anti-symmetric tensor densities  are scalar or vector densities.},
 \be
 \po_\xi\LL=D_\si(\LL\xi^\si)=\di_\si(\LL\xi^\si).
 \lb{24} 
 \ee
 Thus, if $ \de=\po_\xi$, (\ref{23}) minus (\ref{24}) must be equal to zero:
\be
 \po_\xi\LL-\di_\si(\LL\xi^\si)=\LL^A\po_\xi y_A+\di_\si\LLL\fr{\di\LL}{\di(D_\si y_A)}\po_\xi y_A - \LL\xi^\si\RRR=0.
 \lb{25}
 \ee 
(\ref{25}) is commonly referred to as Noether's identity. Next we assume that  $ \po_\xi y_A$ has this form
 \be
 \po_\xi y_A=\La_{A\la}\xi^\la +\La_{A\la}^\rho D_\rho\xi^\la.
 \lb{26}
 \ee
 This is the case for tensor fields. 
The $\La$'s   depend on $y_A$ and $D_\la y_A$ and are   tensors. Inserting (\ref{26})  into $\LL^A\po_\xi y_A$,  we   see from (\ref{25}) that this expression,  which we designate by $X$,  is   necessarily of the following form:
 \be
 X\eq\LL^A\po_\xi y_A=\LL^A\La_{A\la}\xi^\la+\LL^A\La_{A\la}^\rho D_\rho\xi^\la=\di_\si\lll     J^\si_\la\xi^\la+  U^{\si\rho}_\la D_\rho\xi^\la      \rrr.
 \lb{27}
 \ee
 The  $J$'s and  $U$'s are   components of tensor  densities.  A derivation by part of the second term after the first equality sign leads to what has been called since the late 1940's generalized Bianchi identities because this is what they are if $y_A$ is the gravitational field:
 \be
 \LL^A\La_{A\la}-\di_\rho(\LL^A\La^\rho_{A\la})=0.
\lb{28}
\ee

All this has been derived in innumerable papers but has been re-derived here    to clarify our notations.
 \vs
 {\it (ii) The   ``cascade identities" of Julia and Silva}
 \vs
In their paper of 1998 \cite{JS1}, $\xi^\la$ is replaced by
\be
\xi^\la=\xi_0^\la\ep,
\lb{29}
\ee 
where   $\xi^\la_0(x^\mu)$ is  regarded as ``fixed"   for a moment and $\ep(x^\mu)$ is an arbitrary scalar function. Inserting (\ref{29}) into    (\ref{27}) and expanding in terms of $\ep, \di_\rho\ep$ and $D_{\rho\si}\ep$,  one obtains for the \lhs of (\ref{27}), an expression of this form
\be
  X=  X_0\ep+  W^\rho_0 \di_\rho\ep ~~~{\rm where}~~~  W_0^\rho\eq \LL^A\La_{A\la}^\rho \xi^\la_0.
\lb{210}
\ee
$X_0$ is $X$ in which $\xi^\la$ has been replaced by $\xi^\la_0$.
In the \rhs of (\ref{27}) we have a divergence of  
\be
 J^\si_\la\xi^\la+U^{\si\rho}_\la D_\rho\xi^\la=J^\si_0\ep+U^{\si\rho}_0\di_\rho\ep ~~~{\rm with}~~~J_0^\si\eq J^\si_\la\xi^\la_0+  U^{\si\rho}_\la D_\rho\xi^\la_0      ~~~{\rm and}~~~ U^{\si\rho}_0\eq U^{\si\rho}_\la\xi^\la_0.
\lb{211} 
\ee 
But following (\ref{27}), (\ref{210}) is equal to the divergence of (\ref{211}) and since $\xi^\la_0$ is arbitrary the identity   also holds if we replace $\xi^\la_0$ by any  $\xi^\la$ (we remove then the indice $0$). Thus,
\be
  X\ep+  W^\rho \di_\rho\ep=\di_\si\lll  J^\si\ep+  U^{\si\rho}\di_\rho\ep     \rrr= \di_\si  J^\si\ep+\lll J^\rho+D_\si U^{\si\rho}   \rrr\di_\rho\ep+U^{(\rs)}D_{\rs}\ep.
\lb{212}
\ee
  $\ep $ being arbitrary,   one can identify the coefficients of $\ep, \di_\rh\ep$ and $D_{\rs}\ep$ of both sides:
 \be
 X=\di_\rho    J^\rho~~,~~  W^\rho=  J^\rho+\di_\si  U^{\si\rho}~~{\rm and }~~  U^{(\rho\si)}=0.
   \lb{213}
 \ee 
 The interesting point is that $U^{\rs}$ is anti-symmetrical, $U^{\rho\si}=U^{[\rho\si]}$, and   the identities may be rewritten
 \be
  X=\di_\rho    J^\rho~~,~~  J^\rho=  W^\rho+\di_\si  U^{\rho\si}.
  \lb{214}
 \ee
 
 If $\LL^A=0$, then $X=0$,  $W^\rho=0$ and the vector density  $J^\rho$ is     the divergence of an antisymmetric tensor density:
 \be
 J^\rho=\di_\si U^{\rs}~~~~~~~~~~~~,~~~~~~~~~~({\rm only ~if~}\LL^A=0).
 \lb{215}
 \ee 
 (\ref{215}) will not be used later. 
   \vs
 {\it (iii) Functional equation for a superpotential}
 \vs
 Now first for the motivation. Consider the integral, over a volume   $V$ with boundary  $S$,  of $J^\rho$ as given by  (\ref{214}):   
  \be 
 \int_V  J^\rho dV_\rho=\int_VW^\rho dV_\rho+ \oint_S  U^{\rs}dS_{\rs}.
 \lb{216}
 \ee  
In particular, let the volume be a spacelike hypersurface, say $x^0=0$. (\ref{216}) can then be written
\be
\int_V  J^0 d^3x=\int_VW^0 d^3x +\oint_S  U^{0k}dS_k.
\lb{217}
\ee
 Let further  $\xi^\la$  be a timelike vector  which on the boundary, at spatial  or null infinity, is associated with time translations in a flat background. The timelike  field exists in the whole \sss (no black holes). In coordinates in which $\xi^\la=\{1,0,0,0\}$ the volume integral becomes
\be
\int_V  J^0 d^3x=\int_V\LL^A\La_{A0}^0   d^3x +\oint_S  U^{0k}dS_k.
\lb{218}
\ee
In general relativity, (\ref{218}) is  the Hamiltonian:
\be
\int_V  J^0 d^3x=\int_V \sqg (G^0_0-\ka T^0_0) d^3x +\oint_S  U^{0k}dS_k~~~{\rm with}~~~G^\rho_\si= R^\rho_\si-\tha\de^\rho_\si R,
\lb{219}
\ee
  $R_{\rs}$ is the Ricci tensor and $R$ the scalar curvature while $T^\rho_\si$ is the energy momentum of the source of gravity and $\ka$  is the usual coupling constant\footnote{About $\ka$ see (\ref{47}) below and the corresponding footnote.}.
 When the field equations are satisfied,  the Hamiltonian is equal to the total mass-energy:
\be
Mc^2= \oint_S  U^{\rs}dS_{\rho\si}=\oint_S  U^{0k}dS_k~~~~~~~~~~~~,~~~~~~~~~~({\rm only ~if~}\LL^A=0).
\lb{220}
\ee
  $U^{\rs}$ is thus the superpotential by definition. 

The Hamiltonian  field \eeee are obtained by applying the variational principle   to the Hamiltonian. 
Regge and Teitelboim \cite{RT} brought attention to the fact that {\it if boundary conditions are    given to begin with}, the variation of  the Hamiltonian should have no boundary terms like in classical mechanics. Hamiltonians have no time derivatives. Silva's prescription is a covariant expression of that remark. So here is how it goes.

Consider   the variation of (\ref{216}) due to arbitrary variations  of $y_A$; it can be written
 \be
 \de \lll  \int_V  J^\rho dV_\rho\rrr =\int_V\fr{\de W^\rho}{\de y_A} \de y_A dV_\rho
+ \oint_S\lll W^{\rs} +  \de U^{\rs}\rrr dS_{\rho\si}~~{\rm where}~~W^{\rs}\eq \fr{\di W^\rho}{\di (D_\si y_A)}\de y_A.
 \lb{221}
 \ee
 The condition that no boundary term appear in (\ref{221}) provides a functional differential equation for the superpotential:
 \be
  \de  U^{\rs}= -   W^{\rs}  ~~~{\rm so~that}~~~\de \lll\int_V  J^\rho dV_\rho\rrr =\int_V  \fr{\de  W^\rho}{\de y_A}  \de y_AdV_\rho. 
  \lb{222}
 \ee
   Notice that the surface integral in (\ref{221}) and equation (\ref{222}) are only correct if $W^{(\rho\si)}=0$.
That this is indeed true can be seen as follows. Consider the variational derivatives of $X$,   which according to (\ref{213}) on the left, must be equal to zero:
\be
\fr{\de X}{\de y_A}= \fr{\di X}{\di y_A} - D_\rho\lll  \fr{\di X}{\di(D_\rho y_A)} \rrr =0~~~{\rm where}~~~X= \LL^A\La_{A\la}\xi^\la+\LL^A\La^\rho_{A\la} D_\rho\xi^\la.
\lb{223}
\ee 
 
 Now in (\ref{223}) replace $\xi^\la$ by $\xi^\la_0\ep$, expand as a polynomial in $\ep$, $\di_\rho\ep$ and $D_{\rs}\ep$  and equate the factors of $\ep$, $\di_\rho\ep$ and $D_{\rs}\ep$  to zero. Then remove the indice $0$ because $\xi^\la_0$ is arbitrary. The factor of $D_{\rs}\ep$ is $W^{(\rs)}$. Thus $W^{\rs}$ is indeed  antisymmetrical.

Next suppose that $ \us{0}{U}^{\rs}$ is a solution of   equation (\ref{222}) on the left. Then, $\us{0}{U}^{\rs}+C^{\rs}$, where $C^{\rs}(x^\la)$ is an arbitrary function independent of $y_A$, is also a solution of (\ref{222}). When the superpotential relates, in particular to mass-energy,   $C^{\rs}$ defines its   ``zero" point. As it was done in   \cite{KBL} and also in \cite {JS2}, we take $C^{\rs}= - \Bar{ \us{0}{U}^{\rs}}$ which is the superpotential of the background. The superpotential of the background is obtained from $\us{0}{U}^{\rs}$ by equating to zero the source terms\footnote{ For instance, in the \Sc spacetime we take $m=0$ in whichever coordinates the metric is written.  Notice that if  \ssss   are asymptotically flat, $ \Bar{\us{0}{U}^{\rs}}dS_{\rs}=0$ while if they are anti-de Sitter, $\Bar{ \us{0}{U}^{\rs}}dS_{\rs}=\in$ but  in both cases ${U}^{\rs}dS_{\rs}$ is bounded as can be seen in the example considered at the end of the last Section.}. The final solution is then given by
$U^{\rs}=\us{0} {U}^{\rs} - \Bar{ \us{0}{U}^{\rs}}$. 
If, however, we   use a \bbb from the very beginning, as we have done here and shall do in our examples, then $\Bar{ \us{0}{U}^{\rs}}\eq 0$ and $\us{0} {U}^{\rs}=U^{\rs}$.
So much about   the works of Silva and that of Julia and Silva. For completeness let us   mention a work of Fatibene, Ferraris and Francaviglia \cite{FFF} which deals with relative conservation laws.
   \sect{Higher order derivatives of \bm$\xi$\ubm}
 
  \vs
  The method described so far   is not applicable to  the Palatini formulation of \GR because the Lie derivatives of the connection coefficients contain second order derivatives of    $\xi^\la$. So, we first    generalize the prescription of Silva to the case where Lie derivatives contain second order derivatives of  $\xi^\la$ .
  \vs
 {\it (i) Basic elements}
 \vs
 Equations (\ref{21}) to (\ref{25}) remain valid here but instead of  (\ref{26}) we   assume that
 \be
 \po_\xi y_A=\La_{A\la}\xi^\la +\La_{A\la}^\rho D_\rho\xi^\la+\La_{A\la}^{\rho\si} D_{(\rho\si)}\xi^\la~~~\Ra~~~\La_{A\la}^{\rho\si} =\La_{A\la}^{\si\rho}. 
 \lb{31}
 \ee
  (\ref{27})  has now this form   
\be
 X= \LL^A\po_\xi y_A=  \di_\si\lll J^\si_\la\xi^\la+\td U_\la^{\si\rho}D_\rho\xi^\la+V_\la^{\si\mn}D_{(\mn)}\xi^\la  \rrr. 
 \lb{32}
 \ee
Set again, like in (\ref{29}),     $\xi^\la= \xi^\la_0\ep$. The left hand side of (\ref{32}) can then be written like this, thanks to (\ref{31}), 
\be
X=X_0\ep+W^\rho_0\di_\rho\ep+Y^{\rs}_0D_{\rs}\ep~~{\rm with}~~{}W_0^\rho\eq \LL^A\La_{A\la}^\rho \xi_0^\la+ 2\La_{A\la}^{\rho\si} D_\si \xi_0^\la ~~{\rm and}   ~~  Y_0^{\rs}\eq \LL^A\La_{A\la}^{\rho\si} \xi_0^\la.
\lb{33}
\ee 
As before, the indice $0$ indicates that  $\xi^\la_0$ is there instead of  $\xi^\la$. 
 Thus (\ref{32}) takes the following form
 \be
X_0\ep+W^\rho_0\di_\rho\ep+Y^{\rs}_0D_{\rs}\ep=  \di_\si\lll J^\si_0\ep+\td    U_0^{\si\rho}\di_\rho\ep +V_0^{\si\mn}D_{\mn}\ep  \rrr~~~{\rm where}~~~  V_0^{\si\mn}=V_0^{\si\nu\mu}.
\lb{34}   
 \ee
  Since (\ref{34}) holds for any $\xi^\la_0$, it also holds for any $\xi^\la$ and 
 \ba
 X\ep+W^\rho \di_\rho\ep+Y^{\rs}D_{\rs}\ep&=& (\di_\si J^\si)\ep+( J^\rho+ D_\si \td U^{\si\rho})\di_\rho\ep \nn
 \\&&~~~~~~~~+(\td U^{(\rho\si)}+D_\la V^{\la\rho\si})D_{\rs}\ep+V^{\la\rho\si}D_{\la\rho\si}\ep. 
 \lb{35}
 \ea
 
 {\it (ii) Cascade \eeee}
 \vs
 In local Minkowski coordinates,   $D_{\la\rho\si}\ep=\di_{\la\rho\si}\ep$ and its factor, $V^{(\la\rho\si)}$   in the \rhs of (\ref{35}), must be zero because there is no similar factor on the \lhs and $\ep$ is arbitrary. Thus,   we must have both
 \be
V^{\la\rho\si} = V^{\la\si\rho} ~~~{\rm and}~~~ V^{(\la\rs)}\eq\ts{\fr{1}{3}}\lll   V^{\la\rho\si} +V^{\rho\si\la} +V^{\si \la \rho} \rrr=0.
 \lb{36}
 \ee
Multiplying (\ref{36}) by $D_{\la\rho\si}\ep$, we get
\be
\lll   V^{\la\rho\si} +V^{\rho\si\la} +V^{\si \la \rho} \rrr D_{\la\rho\si}\ep=0.
\lb{37}
\ee
With (\ref{36}), one finds that  (\ref{37}) can equally be written as follows:
\be
V^{\la\rho\si}  D_{\la\rho\si}\ep = -\ts{\fr{2}{3}}V^{\mn\la}R^\rho_{~\la\mn}\di_\rho
\ep,
\lb{38} 
\ee
and with (\ref{38}), we can now rewrite  (\ref{35}) like this 
\be
X\ep+W^\rho \di_\rho\ep+Y^{\rs} D_{\rs}\ep=  (\di_\si J^\si)\ep+( J^\rho+ D_\si \td U^{\si\rho}  -\ts{\fr{2}{3}}V^{\mn\la}R^\rho_{~\la\mn})\di_\rho\ep +( \td U^{(\rho\si)} +D_\la V^{\la\rho\si}) D_{\rs}\ep.
\lb{39}
\ee
This identity holds for any $\ep$. Thus the factors of $\ep, \di_\rho\ep$ and $D_{\rs}\ep$ from both sides of the equality must be identical, i.e.
\be
X=\di_\si J^\si~~~,~~~W^\rho=J^\rho+ D_\si \td U^{\si\rho}  -\ts{\fr{2}{3}}V^{\mn\la}R^\rho_{~\la\mn}~~~{\rm and}~~~Y^{\rs}=\td U^{(\rs)}+D_\la V^{\la\rho\si}.
\lb{310}
\ee
These cascade \eeee  are   similar to (\ref{213}).
We shall now reduce this set to one that looks exactly like (\ref{213}).
 
\vs
{\it (iii) An equivalent set of identities}
\vs
We start from $W^\rho$ as given in (\ref{310}) in which we replace   $ \td U^{\si\rho}$ by $ ( -  \td U^{[\rs]}+\td U^{(\rs)})$:
\be
{}W^\rho=J^\rho - \di_\si \td U^{[\rho\si]}+ D_\si \td U^{(\si\rho)}   -\ts{\fr{2}{3}}V^{\mn\la}R^\rho_{~\la\mn}.
\lb{311}
\ee
Then  we replace $D_\si\td U^{(\si\rho)} $ by its value deduced from $Y^{\rho\si}$, given in  the \rhs of (\ref{310}), and we get for (\ref{311})
\be
{}W^\rho=J^\rho - \di_\si \td U^{[\rho\si]}+ D_\si Y^{\si\rho} -\lll D_{\si\la}V^{\la\rs}  +\ts{\fr{2}{3}}V^{\mn\la}R^\rho_{~\la\mn}\rrr.
\lb{312}
\ee
However, thanks to the symmetries of $V^{\la\rho\si}$ we readily find that
  \be
  D_{\si\la}  V^{\la\rho\si} + \ts{\fr{2}{3}}V^{\mn\la}R^\rho_{~\la\mn}= -  \ts{\fr{2}{3}} \di_\si    \lll D_\la     V^{[\rho\si]\la}  \rrr.  
\lb{313}
 \ee
 Thus instead of  (\ref{312})  we may also write
 \be
{}W^\rho=J^\rho -\di_\si  U^{\rho\si}+ D_\si Y^{\rs} ~~~{\rm where}~~~ U^{\rho\si}\eq  \td U^{[\rs]} -  \ts{\fr{2}{3}} D_\la     V^{[\rho\si]\la}= - U^{\si\rho}.    
\lb{314}
 \ee
  The tensor $U^{\rs}$ is anti-symmetrical. Thus  (\ref{314}) provides identities similar to  (\ref{213}): 
 \be
 X=\di_\rho J^\rho  ~~~,~~~ J^\rho =*W^\rho +  \di_\si U^{\rho\si} ~~~{\rm where}~~~*W^\rho\eq W^\rho - D_\si Y^{\rs}.
 \lb{315}
 \ee  
and the corresponding  generalized Bianchi identities are
\be
 \LL^A\La_{A\la}- D_\rho(\LL^A\La^\rho_{A\la})+D_{\rs}(\LL^A\La^{\rs}_{A\la})=0. 
\lb{316}
\ee

{\it (iv) More identities}
\vs 
Consider now (\ref{33}) which we rewrite like this:
\be
X =X_0+W^\rho_0\di_\rho\ep+Y^{\rs}_0D_{\rs}\ep= X_0+\td W^\rho_0\di_\rho\ep+Y^{\rs}_0\di_{\rs}\ep ~~{\rm where}~~ \tW^\rho\eq W^\rho -  Y^{\mn}\Ga^\rho_{\mn}.
\lb{317}
\ee
Since $X$ is a divergence we have 
\be
\fr{\de X}{\de y_A}=\fr{\di X}{\di y_A} - D_\si\!\lll\fr{\di X}{\di (D_\si y_A)} \rrr=0.
\lb{318}
\ee
Inserting (\ref{317}) into (\ref{318}), identifying the coefficients of $\ep, \di_\rho\ep, \di_{\rs}\ep$ and $\di_{\la\rs}\ep$ to zero, removing the indice $0$, because the resulting identities are valid for any   $\xi^\la$, one   obtains       four  identities, two of which do not involve $X$: 
\be
  \fr{\de Y^{\rs}}{\de y_A}\de y_A=\tW^{(\rs)} \Ra
  \tW^{\rs} \eq \fr{\di \tW^{\rho}}{\di (D_{\si} y_A)}\de y_A ~~
{\rm and}~~{Y^{(\la\rs)}}=0 \Ra  Y^{\la\rs}\eq\fr{\di Y^{\rs}}{\di(D_\la y_A)}\de y_A.
 \lb{319}
\ee
We notice that $Y^{\la\rho\si}$ has the same symmetries as $V^{\la\rho\si}$. Therefore (\ref{313}) holds for $Y^{\la\rho\si}$ except that   in (\ref{317}) we have ordinary derivatives as if we were in a flat space and therefore
\be
  \di_{\si\la}(Y^{\la\rho\si} )= -  \ts{\fr{2}{3}}\di_\si    \lll \di_\la  Y^{[\rho\si]\la}  \rrr.   
 \lb{320}
\ee
 
{\it (v) The \sup \eee}
\vs 
We now consider the variation of $J^\rho$ as given in (\ref{315}), using $\tW^\rho$ as defined in  (\ref{317})
\be
  \de J^\rho= \de(W^\rho - D_\si Y^{\rs})+\di_\si \de U^{\rs}= \de \tW^\rho - \di_\si \de Y^{\rs}+\di_\si \de U^{\rs}.
\lb{321}
\ee
Notice that    $\tW^\rho$  like $Y^{\rho\si}$  contains at most first order derivatives of $y_A$. Therefore, we may write (\ref{321}), making use of  (\ref{317}) as follows:
\be
\de J^\rho= \fr{\de \tW^\rho}{\de y_A} \de y_A+ \di_\si     \LLL  \tW^{\rs}-\fr{\di Y^{\rs}}{\di y_A} \de y_A -  \lll\fr{\di Y^{\rs}}{\di(D_\la y_A)}\rrr \di_\la \de y_A +   \de U^{\rs}\RRR.
\lb{322}
\ee
\setlength{\baselineskip}{20pt plus2pt}
 With   (\ref{319}), the sum of the first two  terms in parenthesis may be written
  \be
  \tW^{\rs} - \fr{\di Y^{\rs}}{\di y_A} \de y_A=  \tW^{[\rs]} - \di_\la \lll\fr{\di Y^{\rs}}{\di(D_\la y_A)}\rrr \de y_A,
  \lb{323}
 \ee
and therefore  $\de J^\rho$ can also be written 
 \be
 \de J^\rho= \fr{\de \tW^\rho}{\de y_A} \de y_A+ \di_\si[\tW^{[\rs]} - \di_\la (Y^{\la\rs})+   \de U^{\rs}]. 
   \lb{324}
 \ee
 But using (\ref{320}) and the definition of  $\tW^\rho$  in (\ref{317}) we find that
 \normalsize
 \be
 \di_\si \tW^{[\rs]} -\di_{\si\la} (Y^{\la\rs})= \di_\si[W^{[\rs]} +  {\ts{\fr{2}{3}}}D_\la (Y^{[\rs]\la})],  
 \lb{325}
  \ee
  $W^{\rs}$ has been defined in (\ref{221}).
  So
  \be
   \de J^\rho= \fr{\de \tW^\rho}{\de y_A} \de y_A+ \di_\si   [ W^{[\rs]} +  \ts{\fr{2}{3}}D_\la (Y^{[\rs]\la})+   \de U^{\rs}]. 
   \lb{326}
\ee
\setlength{\baselineskip}{20pt plus2pt}
But, see (\ref{315}) and (\ref{317}), 
\be
\tW^\rho=W^\rho - Y^{\mn}\Ga^\rho_{~\mn}=*W^\rho + \di_\si Y^{\rs} ~~~{\rm so~that}~~~~\fr{\de \tW^\rho}{\de y_A} =\fr{\de *W^\rho}{\de y_A}. 
\lb{327}
\ee
 Thus the differential \eee for the superpotential   is   now
 \be
   \de U^{\rs}= - W^{[\rs]} -  {\ts{\fr{2}{3}}}D_\la (Y^{[\rs]\la})~~~{\rm and}~~~ \de \lll\int_V J^\rho dV_\rho\rrr= \int_V \fr{\de *W^\rho}{\de y_A} \de y_A dV_\rho.  
\lb{328}
 \ee

Putting   together the various elements defined along the way, we have    an equation for the superpotential - the first equality in (\ref{328}) - that looks like this:
\be
  \de U^{\rs}= -  \fr{\di W^{[\rho}}{\di (D_{\si]}y_A ) } \de y_A + { \ts{\fr{2}{3}} }D_\la\lll    \fr{\di Y^{\la[\rho}}{\di(D_{\si]}y_A )} \de y_A \rrr   ,
   \lb{329}
\ee
in which
\be
W^\rho= \LL^A\La_{A\la}^\rho \xi^\la+  2\LL^A\La^{\rs}_{A\la}D_\si\xi^\la~~~{\rm and}~~~Y^{\rs}=\LL^A \La^{\rs}_{A\la}\xi^\la.
\lb{330}
\ee
Equations (\ref{329}) and (\ref{330}), together  with appropriate boundary         conditions, are all we need to calculate the superpotential in the following example.  
 \sect{Application to variational derivatives of Lovelock Lagrangians in the  Palatini representation }
 
 \vs
 {\it (i) Lagrangian, variational derivatives and boundary conditions}
 \vs
  In the Palatini formulation we take the inverse metric components $g^{\mn}$ and the connection coefficients $\Ga^\la_{~\rs}=\Ga^\la_{~\si\rh}$ as independent field components.   The   curvature tensor   does not have all the symmetries of the Riemannian one. To avoid any confusion we shall denote the curvature tensor, like in Eisenhart \cite{Ei}, by $B^\la_{~\nu\rs}$ and reserve the usual notation of the curvature tensor $R^\la_{~\nu\rs}$ for when the connection coefficients are Christoffel symbols. $B^\la_{~\nu\rs}$  is antisymmetrical in the last two indices only, $B^\la_{~\nu\rs}= - B^\la_{~\nu\si\rho}$. Additional symmetry properties, similar to those of the Riemann curvature tensor, are:
 \be
 B^\la_{~(\nu\rs)}=0~,~B^\la_{~\nu(\rs;\tau)}=0~~{\rm and}~~D_\la B^\la_{~\nu\rs}=D_{\rh} B_{\nu\si} - D_{\si} B_{\nu\rh} ~~{\rm where}~~B_{\si\nu}=B^\la_{~\si\la\nu}\ne B_{\nu\si}.
 \lb{41}
 \ee
Here a semi-column means covariant differentiation. The curvature tensor itself is given by
\be
B^\la_{~\nu\rs}= 2(\di_{[\rho}\Ga^\la_{~\si]\nu} + \Ga^\la_{~\eta[\rho}\Ga^\eta_{~\si]\nu}). 
\lb{42}
\ee
In accordance with \cite{KBL}, we introduce a \bbb whose metric components are $\Bar g_{\mn}$.  Instead of $\Ga$'s we introduce   
\be
\De^\la_{~\rs}\eq\Ga^\la_{~\rs} - \Bar{\Ga}^\la_{~\rs}=\tha g^{\la\eta}(\BD_\rh g_{\eta\si}+\BD_\si g_{\eta\rh} - \BD_\eta g_{\rs}).
\lb{43}
\ee
The bars like on $\BD$'s refer to the background. 
$\De^\la_{~\rs}$ is a tensor. The curvature tensor can be written in terms of these $\De$'s:
\be
B^\la_{~\nu\rs}= 2(\BD_{[\rho}\De^\la_{~\si]\nu} + \De^\la_{~\eta[\rho}\De^\eta_{~\si]\nu})+\Bar{B}^\la_{~\nu\rs}. 
\lb{44}
\ee

  As boundary conditions we want to impose the value of the metric components on the boundary $S$, i.e.
  \be
  g_{\mn}|_S=\Bg_{\mn}.
  \lb{45}
  \ee
We shall also demand that on the boundary the $\Ga$'s be   Christoffel symbols i.e. that
    \be
   D_\la g^{\mn}|_S=0~~~{\rm and~ thus}~~~B^\la_{~\nu\rs}|_S= R^\la_{~\nu\rs}. 
   \lb{46}
    \ee
 The action of the gravitational field is of the form
 \be
 A=\2k\int \LL(g_{\mn}, B^\la_{~\nu\rs})d^Nx~~{\rm where}~~\ka=\fr{2S_{N-2}G_N}{c^4};
 \lb{47}
 \ee   
 $\LL$ is a Lovelock Lagrangian  \cite{Lo},  $N$ is the dimension of spacetime. The coupling constant $\ka$ is normalized like in \cite{AD}, except for a factor 2:  $S_{N-2}$ is the surface of a sphere of dimension $N-2$ and $G_N$ the gravitational coupling constant.\footnote{For $N=4$, $S_2=4\pi$ and $\ka = \fr{8\pi G}{c^4}$ in which $G$ is Newton's gravitational constant.} 
 
 Contrary to our assumption in Sections 2 and 3 that \sss has a passive role here it is part of the game: the metric components and the connection coefficients {\it are} the field components.  Nonetheless,   Section 3 is applicable to variational derivatives of   Lovelock Lagrangians because:  (1) the connection coefficients enter into tensorial combinations,   (2) the variational derivatives as well as Lie derivatives of the connection coefficients are tensors as well and (3)   on the boundary the connection coefficients are Christoffel symbols.
 
 The variation of the action has this form
\be
 \de A=  \int (\LL_{\mn}\de g^{\mn} + \LL^{~\rs}_\la\de\De^\la_{~\rs})d^Nx +   \oint_S \DD^\mu dS_\mu.
 \lb{48}
\ee
$\LL_{\mn}$ and $\LL_\la^{~\rs}$ are variational derivatives.  
In applying   the variational principle to the action of the gravitational field plus its sources, ($\de A+\de A_{\rm sources}=0$), $\LL_{\mn}$      is defined by the sources but for minimally coupled matter,  $ \LL^{~\rs}_\la=0$. This last equation is linear and homogeneous in $D_\la g^{\mn}$ and contains no derivatives of $B^\la_{~\nu\rs}$. This is the particularity of Lovelock Lagrangians. As a consequence, and as shown by Exirifart and Sheikh-Jabbari \cite{ES}, $D_\la g^{\mn}=0$ will always be a solution of this equation. We shall, of course, not use that solution except to notice that it is in accordance with the bounday condition (\ref{46}). In addition to  $ \LL^{~\rs}_\la=0$,   the variational  principle would impose  some boundary condition i.e., for {\it isolated sources},
\be
\DD^\mu dS_\mu=0.
\lb{49}
\ee
Adding a divergence to the Lagrangian may change this condition but, contrary to what happens in the Hamiltonian formalism, there is no way to get rid of  a condition like (\ref{49}) by adding a divergence to $\LL$. The condition (\ref{49}) may   be satisfied if (\ref{45}), (\ref{46}) hold. Otherwise (\ref{49}) must at least  be compatible with (\ref{45}), (\ref{46}). 
\vs
{\it (ii) The \sup }
\vs
 The Lie derivatives ${\po_\xi y_A}$ are as follows, the second equality is a repeat in terms of $\La$'s like in (\ref{31}) where indices $A$ are surrounded by curly brackets:
 \ba
 \po_\xi    g^{\mn}&=& D_\la   g^{\mn}\xi^\la -   2g^{\eta(\mu} D_{\eta} \xi^{\nu)}=\La^{\{\mn\}}_\la \xi^\la+\La^{\{\mn\}\eta}_\la  D_\eta\xi^\la,
 \lb{410}\\
 \po_\xi \Ga^\eta_{\rs}&=& - R^\eta_{~(\rs)\la}\xi^\la+D_{(\rs)}\xi^\eta=\La^{\{\eta\}}_{{\{\rs\}}\la}\xi^\la+\La^{\{\eta\}\mn}_{{\{\rs\}}\la}D_{(\mn)}\xi^\la.
 \lb{411}
 \ea 
 The only non-zero $\La$'s  appearing in $W^\rho$ and $Y^{\rho\si}$, see (\ref{330}), are  
 \be
 \La^{\{\mn\}\eta}_\la  = -   2g^{\eta(\mu}\de^{\nu)}_\la ~~~{\rm and}~~~\La^{\{\eta\}\mn}_{\{\rs\}\la}=\ts{\fr{1}{2}} (\de_\rh^\mu  \de_\si^\nu  +\de_\rh^\nu  \de_\si^\mu)\de ^\eta_\la. 
 \lb{412}
 \ee
Introducing these $\La$'s into $W^\rho$ and $Y^{\rho\si}$, we obtain
\be
W^\rho= 2(- \LL^\rh_\la \xi^\la+\LL^{~\rs}_\la D_\si\xi^\la)  ~~~{\rm and}~~~Y^{\rs}=\LL^{~\rs}_\la\xi^\la.
\lb{413}
\ee
With  this, the solution of equation (\ref{329}) is absolutely straightforward. Since $g_{\mn}|_S=\Bar g_{\mn}$, we have $\de g_{\mn}=0$ in the equation and if (\ref{46}) and (\ref{47}) hold,   the following equalities must also hold on the boundary $S$:
\be
D_\eta\de g^{\mn}|_S=  - 2\Bg^{\tau(\mu}\de\De^{\nu)}_{~\tau\eta} ~~~{\rm and}~~~\de(D_\eta\xi^\la)|_S=\xi^\tau \de\De^\la_{~\tau\eta}.
\lb{414}
\ee
The equation for $U^{\rs}$ reduces  thus to a linear differential form in $\de\De^\la_{~\mn}$.  
With  (\ref{414})  and $\de g_{\mn}=0$, the equation for the \sup (\ref{329})  may be written
 \be
 \de U^{\rs}= - 4\LLL  \fr{\di\LL^{[\rh}_\tau\xi^\tau}{\di B^\la_{~\mn\si]}}  + \fr{1}{3}\fr{\di \LL_\tau^{~\mu[\rh}\xi^\tau}{\di(D_{\si]}g^{\eta\la})}g^{\eta\nu}\RRR_S \de\De^\la_{\mn}.
\lb{415}
 \ee
 In this expression it is understood that the square brackets $\ts{[\rh}$  and $\ts{\si]}$ mean that the expression is anti-symmetrized in $\ts{\rs}$.
Since the factor of $\de\De^\la_{\mn}$ contains only the metric and the curvature tensor which on the boundary are given thanks to (\ref{45}) and (\ref{46}), provided the $\BGa$'s are derivable at least once, we may write that
 \be
   U^{\rs}= - 4\LLL  \fr{\di\LL^{[\rh}_\tau\xi^\tau}{\di B^\la_{~\mn\si]}}  + \fr{1}{3}\fr{\di \LL_\tau^{~\mu[\rh}\xi^\tau}{\di(D_{\si]}g^{\eta\la})}g^{\eta\nu}\RRR_S  \De^\la_{\mn}.
\lb{416}
 \ee
 Notice that $\Bar U^{\rs}=0$ because $\Bar\De^\la_{~\rs}=0$. 
  $U^{\rs}$ is the \sup for   theories of gravity derived from a Lovelock Lagrangian.
 \sect{Application to \GR in $N$ dimensional \ssss}
\vs
{\it (i) The variational derivatives}
\vs
 Let
\be
B\eq g^{\mn}B_{\mn}=b^{\mn\rs}B_{\mn\rs},
\lb{51}
\ee
where
 \be
 b^{\mn\rs} \eq{g^{\mu[\rh}g^{\si]\nu}}.
 \lb{52}
 \ee  
Then, if a hat like in $\hg_{\mn}$  means multiplication by $\sqg$, 
the ``relative" Einstein-Hilbert action is\footnote{The $1/2\ka$ factor is for convenience.}
\be
A_1=    \int \LL_1 d^Nx\eq   \2k\int(\hat B - \Bar{\hat B})d^Nx. 
\lb{53}
\ee
 The variation of the action,  
 \be
 \de A_1=  \int (\LL_{\mn}\de g^{\mn} + \LL^{~\rs}_\la\de\De^\la_{~\rs})d^Nx - \1k \oint_S \hat b^{\mu\rs}_{~~~~\la}\de\De^\la_{~\rs}dS_\mu,
 \lb{54}
 \ee
 
The variational derivatives,
 \be
 \LL_{\mn}=\fr{\de\LL_1}{\de g^{\mn}}= \2k( \hat B_{(\mn)} - \ha g_{\mn}\hat B) ~~~{\rm and}~~~\LL_\la^{~\rs}=\fr{\de\LL_1}{\de \Ga^\la_{~\rs}}=\2k (  \de^{(\rh}_\la D_\eta  \hg^{\si)\eta} - D_\la \hg^{\rs} ).
\lb{55}
 \ee
  We may and have indeed used $B_{\mn}$ in terms of \bbb derivatives and in terms of $\De$'s so as to keep every term covariant. 
 The boundary conditions that would follow from the Action principle would be $\Bar{\hat b^{\mu\rs}_{~~~~\la}}\de\De^\la_{~\rs}dS_\mu=0$. Since neither $\Bar{\hat b^{\mu\rs}_{~~~~\la}}=0$ nor $\de\De^\la_{~\rs}dS_\mu=0$ are, in general, compatible with (\ref{45}) and (\ref{46}), we shall add a  divergence and take the following  new action\footnote{This is not the only surface term possible which leads to acceptable boundary conditions. We might also add $\1k\int_S\Bar{\hat b^{\mu\rs}}_{\la}\De^\la_{~\rs}dS_\mu$ instead to $A_1$.}
 \be
 A'_1=A_1+\1k\oint_S\hat b^{\mu\rs}_{~~~~\la}\De^\la_{~\rs}dS_\mu.
 \lb{56}
 \ee  
 The variation of this action is 
 \be
  \de A'_1=  \int (\LL_{\mn}\de g^{\mn} + \LL^{~\rs}_\la\de\De^\la_{~\rs})d^Nx + \1k \oint_S \de(\hat b^{\mu\rs}_{~~~~\la})\De^\la_{~\rs}dS_\mu.  
  \lb{57}
  \ee
    The   boundary conditions may now be $\de(\hat b^{\mu\rs}_{~~~~\la})=0$ or $b^{\mn\rs}|_S=\Bg^{\mu[\rh}\Bg^{\si]\nu}$ and these conditions are compatible with  (\ref{45}). 
    \vs
{\it (ii) The \sup}
\vs
The superpotential (\ref{416}) in this case is  simple to calculate:
  \be
 U_1^{\rs}=  \fr{3}{\ka}\hat\xi^{(\mu}   b^{\rs)\nu}_{~~~~~\la}\big{|}_S\De^\la_{~\mn}=\fr{3}{\ka}\hat\xi^{(\la} b^{\rs)\mn}\big{|}_S\BD_\mu g_{\nu\la}.
 \lb{58}
 \ee
 The last equality follows from (\ref{43}).
  $U_1^{\rs}$ is, as expected, the KBL superpotential.  This may not be apparent because 
the  KBL \sup is more often  written {\it locally} like this \cite{KBL}:
\be
 U_1^{\rho\si}  =\1k\lll  D^{[\rho}  \hat\xi^{\si]} - \Bar{ D^{[\rho} \hat \xi^{\si]}}\rrr +\1k\hat\xi^{[\rho}k^{\si]} ~~{\rm where}~~ k^{\si}\eq\hg^{\si\mu}\De^\nu_{\mn} - \hg^{\mn}\De^\si_{\mn}.    
\lb{59}
\ee
  To see that (\ref{59}) is   the same as  (\ref{58}) we notice that   $k^\si$ can be written  as follows:
 \be
 k^\si=2b^{\si\mn}_{~~~~\la}\De^\la_{~\mn}~~{\rm and}~~ D^{[\rho}  \hat\xi^{\si]} - \Bar{ D^{[\rho} \hat \xi^{\si]}}=(\hg^{\mu[\rh}-\Bar{\hg^{\mu[\rh}})\BD_\mu\xi^{\si]}+\hg^{\mu[\rh}\De^{\si]}_{~\mn}\xi^\nu.
\lb{510}
 \ee
 But since $g_{\mn}|_S=\Bar{g}_{\mn}$,  the first term on the \rhs is zero. So,   taking (\ref{59})  into account   and using the expression  for the tensor \bm{$b~$}\ubm defined  in (\ref{52}),  one easily finds   that (\ref{510}) is  the same as (\ref{58}).

  One may also write (\ref{58}) like this:
 \be
U_1^{\rho\si}  =\1k\xi_\mu \BD_\nu (\Bg^{\mu[\rh}\hg^{\si]\nu} - \Bg^{\nu[\rh}\hg^{\si]\mu}).
\lb{512} 
 \ee
It appears   in this form, on a sphere at   spatial infinity, in \cite{AD} and \cite{DT2} with $\hg^{\mn}$ to leading order in $1/r$.   
 The   physical properties of the KBL \sup in \GR are summarized in \cite{JS2}. It is worth noting that the superpotential is valid on any $S$, whether at spatial {\it or} at null infinity. 
 \sect{Application to Gauss-Bonnet theories in the Palatini representation}
  
 \vs
 {\it (i) Lagrangian, variational derivatives  and boundary conditions }
 \vs
 Gauss-Bonnet theories of gravity\footnote{Gauss-Bonnet theories of gravity are rarely considered alone. What is usually used is   Einstein's theory in $N$ dimensions to which a Gauss-Bonnet term is added. This is then called an Einstein-Gauss-Bonnet theory   which we consider briefly below.} in a Palatini formulation    apply to \ssss with  more that $4$ dimensions.  Another difference with Einstein's theory of gravitation lies in the relation between the metric and the connection coefficient. In Einstein's theory $\LL_\la^{~\rs}=0$ implies   $D_\la g^{\mn}=0$. In Gauss-Bonnet theories  $D_\la g^{\mn}=0$ is always a solution of $\LL_\la^{~\rs}=0$.  
 
   Equations (\ref{41}) to (\ref{46}) and (\ref{410}) to (\ref{416}) hold in Gauss-Bonnet theories.   
The variational derivatives are derived from a Lovelock Lagrangian of order two which we took from Jacobson and Myers \cite{JM}. Define 
\be
\BB\eq  g^{\mn}g^{\rs}( B^\eta_{~\mu\la\rh}B^\la_{~\si\eta\nu} - B^\eta_{~\mn\la}B^\la_{~\rs\eta} - 2B^\la_{~\rh\mu\si}B_{\nu\la}+B_{\mn}B_{\rs} - B_{\rh\mu}B_{\nu\si}).
\lb{61}
\ee
Then, the  relative action with respect to the background,
\be
A_2=\int \LL_2d^Nx\eq\fr{1}{4\ka}\int (\hat \BB - \Bar{\hat\BB})d^Nx.
\lb{62}
\ee
The variation of the action is of this form:
 \be
 \de A_2=  \int (\LL_{\mn}\de g^{\mn} + \LL^{~\rs}_\la\de\De^\la_{~\rs})d^Nx -   \fr{1}{\ka}\oint_S \hat P^{\mu\rs}_{~~~~\la}\de\De^\la_{~\rs}dS_\mu;
 \lb{63}
 \ee
in this 
the variational derivatives with respect to the metric components are as follows:
\be
\LL_{\mn} = \fr{\de\LL_2}{\de g^{\mn}}=-\tha\LL_2\hg_{\mn}+\sqg \tLL_{\mn}
\lb{64}
\ee
in which
\be
2\ka \tLL_{\mn}= g^{\rs}( B^\eta_{~\rh\la(\mu}B^\la_{~\nu)\eta\si} + B^\eta_{~(\mn)\rh}B_{\si\eta} + B_{\rs}B_{(\mn)} +  B^\eta_{~\rs(\mu}B_{\nu)\eta} - B_{\rh(\mu}B_{\nu)\si} - B^\eta_{~(\mn)\la}B^\la_{~\rs\eta}),
 \lb{65}
\ee
while variational derivatives \wrt the connection coefficients are of the form $D[\OO(g)B]$ where $\OO$ depends on the metric components only, the Bianchi identities eliminate the derivatives of   $B$'s and the result is of the form $D[\OO(g)]B$; explicitly:
\ba
2 \ka\LL_\la^{~\rs}\!\!\!\!\!&=&\!\!\!2\ka \fr{\de\LL_2}{\de \Ga^\la_{\rs}}=\ts{\fr{3}{2}}D_\eta[\hg^{\mu(\rh}\de^\tau_\la g^{\nu)\si}+\hg^{\mu(\si}\de^\tau_\la g^{\nu)\rh}]B^\eta_{~\mn\tau}+ 3D_\eta(\hg^{\mu(\nu}\de^\tau_\la g^{\eta)(\rh}) B^{\si)}_{~\mn\tau}\nn\\
&&~~~~~~~~~~~+\ts{\fr{3}{2}} D_\eta(\de_\la^{(\nu}\hat b^{\si\eta)\rh\mu}+\de_\la^{(\nu}\hat b^{\rh\eta)\si\mu} )B_{\mn}
\lb{66}
\ea

 The {\it boundary term} introduces a tensor  denoted by  Davis \cite{Da} as \!\bm{ $P~$}\ubm:
 \be
 P^{\mn\rs}|_S\eq (R^{\mn\rs} - 2g^{\mu[\rh} R^{\si]\nu}+ 2g^{\nu[\rh} R^{\si]\mu}   + Rg^{\mu[\rho} g^{\si]\nu}).
 \lb{67}
 \ee
$P^{\mn\rs}$, which has manifestly   the symmetries of the curvature tensor like $b^{\mn\rs}$,     is  also divergenceless: $D_\mu P^{\mn\rs}=0$.
  
 We may take a new action which is indeed compatible with (\ref{45}) and (\ref{46}) by adding a divergence to $\LL_2$:
 \be
A'_2=\int \LL_2 d^Nx +\fr{1}{\ka}\oint_S \hat P^{\mu\rs}_{~~~~\la}\De^\la_{~\rs}dS_\mu, 
\lb{68}
 \ee
 so that
\be
 \de A'_2=  \int (\LL_{\eta\la}\de g^{\eta\la} + \LL^{~\rs}_\la\de\De^\la_{~\rs})d^Nx +  \fr{1}{\ka} \oint_S \de( \hat P^{\mu\rs}_{~~~~\la})\De^\la_{~\rs}dS_\mu.
 \lb{69}
 \ee
 The boundary condition that follows from the variational principle is now  
 $\de \hat P^{\mn\rs} =0$ which is equivalent to (\ref{67}).
  This condition will automatically hold  if   (\ref{45}) and (\ref{46}) are satisfied    and the $\BGa$'s are  derivable at least once.    
   
     \vs
 {\it (ii) The \sup }
 \vs
 The calculation of the \sup $U_2^{\rs}$   given by equation (\ref{416}) is somewhat more complicated    than for Einstein's theory in $N$ dimensions.   MathTensor  and Mathematica were valuable tools to check our   calculations. A   condensed but readable formula for $U^{\rs}_2$ is perhaps this one:  
 \ba
 U^{\rs}_2&=&3[\hat\xi^{(\mu}R^{\rs)\nu}_{~~~~\la}\De^\la_{~\mn}+(\hg^{\mu(\rh}G^\si_\eta\De^{\nu)}_{~\mn}-\hg^{\mu(\si}G^\rh_\eta\De^{\nu)}_{~\mn})\xi^\eta]\nn\\
 &+&\ts{\fr{7}{2}}(\hg^{\mu(\nu}R^{\rs)}_{~~~\la\eta}\xi^\eta+R^{(\rh}_\la\xi^\mu\hg^{\si)\nu} - R^{(\si}_\la\xi^\mu\hg^{\rh)\nu})\De^\la_{~\mn} \nn\\
& -& \ts{\fr{5}{2}}(\De^{(\mu}_{~\mn}R^{\rs)\nu}_{~~~~\eta}\hat\xi^\eta+ \hat R^{\mu(\rh}\De^\si_{\mn~}\xi^{\nu)}  -  \hat R^{\mu(\si}\De^\rh_{\mn~}\xi^{\nu)}), 
\lb{610}
 \ea
the factors of the $\De$'s are, of course,  to be evaluated on the boundary, that is, in terms of the background geometry which is generally much simpler than that of the spacetime itself.

 Deser and   Tekin \cite{DT1}, \cite{DT2} have calculated the mass-energy in generic higher curvature gravity theories, in particular  on anti-de Sitter backgrounds, motivated by the role of these \bbbb in string theory.  
 
 Because  of its importance in string theory and as an opportunity to relate our results with previous  calculations, we consider now what becomes of $U^{\rs}_2$ if the curvature tensor of the background is of the form
\be
\BR_{\mn\rs}=\fr{1}{l^2}(\Bg_{\mu\rh}\Bg_{\si\nu} - \Bg_{\mu\si}\Bg_{\rh\nu})=\fr{2}{l^2}\Bar b_{\mn\rs}.
\lb{611}
\ee
With such a \bbb (\ref{610}) becomes rather simple:\footnote{Formulas (\ref{610}) as well as (\ref{612}) are at variance with the \sup suggested in \cite{DKO}.}
\be
U^{\rs}_2=\fr{1}{l^2}(N-3)(N-4)U^{\rs}_1.
\lb{612}
\ee
We now consider the Einstein-Gauss-Bonnet theory of gravity and follow Deser and   Tekin \cite{DT2}. Our action is written like this
\be
A=\int (\LL_1+a\LL_2)d^Nx+\1k\int 2\La_0(\sqg-\sq{ -\Bg})d^Nx.
\lb{613}
\ee
$a$ is a coupling constant and $\La_0$ contributes to the overall effective cosmological constant $\La$ which 
is related to\footnote{in \cite{DT2}, $l^2$ is defined with the opposite sign so that their $ l^2$ for AdS \ssss is negative.} $l^2$ as follows:
\be
l^2= - \fr{(N-1)(N-2)}{2\La}~~~~,~~~~\La<0.
\lb{614}
\ee
The \sup in this case is   of the form
\be
U^{\rs}=U^{\rs}_1+aU^{\rs}_2=\lll1+a\fr{(N-3)(N-4)}{l^2}\rrr U^{\rs}_1.
\lb{615}
\ee
In the case considered by Deser and   Tekin,    
\be
a= -2 \ka\ga=\fr{ - 2l^2}{(N-3)(N-4)}.
\lb{616}
\ee
Thus, 
\be
U^{\rs}= - U^{\rs}_1.
\lb{617}
\ee
This is indeed their \sup on a sphere at spatial infinity, as can be figured out  from their formula (31) in which  we must set $\al=\bt=0$.  
\sect{A brief summary and some comments}
\vs
{\it (i) First the summary}
\vs
The \sup $U^{\rs}_n$ of a Lovelock Lagrangian of order $n$, $\LL_n$, $n=1,2,3,\cdots$,  or the \sup $U^{\rs}=\sum_n a^nU^{\rs}_n$ of a linear combination  of such Lagrangians $\LL=\sum_n a^n\LL_n$, with coupling constants $a^n$, is given by \eee (4.16). The structure of $U^{\rs}_n$ is rather obvious. $U^{\rs}_n$ is homogeneous of order $(n-1)$ in the curvature tensor {\it of the background} and is  linear homogeneous in $\De^\la_{\rs}=\Ga^\la_{\rs} - \BGa^\la_{\rs}$. 

We calculated $U^{\rs}_1$ which is well known and has been obtained by various authors with different methods. It was, however, useful to show the reader that we    recovered at least well known results. One practical novelty, besides \eee (4.16), is an explicit expression for $U^{\rs}_2$ on arbitrary backgrounds, formula (6.10). Another one is  an  expression for $U^{\rs}_2$, see (6.12),  on anti-de Sitter backgrounds. This later expression provides the same \sup as that found by Deser and Tekin at spatial infinity, which they obtained using a  very different method. The result  gave us further confidence that the procedure of Silva,  extended in Section 3 to Palatini's representation of gravity fields, did indeed   work and that \eee (4.16) was   correct. One might   have calculated $U^{\rs}_3$ and  $U^{\rs}_4$ using   
$\LL_3$ and $\LL_4$ explicitly written in an Appendix of a paper by J. T. Wheeler \cite{Wh}\footnote{See also M\"uller-Hoissen \cite{Mu} who gives $\LL_3$.}.   This would have been hard work with little direct prospect of   applicability. 
The method of Section 3 may of course be applied to variational derivatives of higher order that are not Lovelock Lagrangians, the type of Lagrangians that interest string theorists.
\vs
{\it (ii) Now some comments}
\vs

\nnn a] Regarding $U^{\rs}_1$ as given by \eee (5.8). On a flat background, in Minkowski \coo  and  with \bm $\xi$ \ubm the Killing vector of spacetime translations,  $U^{\rs}_1$ is exactly the \sup that Freud \cite{Fr} found, almost 70 years ago, to calculate mass-energy and total linear momentum. One wonders why Freud did not  calculate the angular momentum on the same occasion.

\nnn b] We emphasized several times that the superpotential holds  at null as well as at spatial infinity and that  $U^{\rs}_1$ at null infinity gives the Bondi mass. We do not know if anybody got interested in radiating fields for Einstein-Gauss-Bonnet theories.  $U^{\rs}_2$ is, of course, the \sup appropriate for calculating the Bondi mass.

\nnn c] We imposed Dirichlet boundary conditions. Julia and Silva showed that imposing Neumann   boundary conditions  lead to Komar's superpotential\footnote{A generalization of Komar's superpotential, that does not need a background, to any type of boundary conditions,  can be found in several papers of Obukhov and Rubilar's ; see for instance \cite{OR}. The role of \bbbb is here replaced by another ingredient, ``generalized" Lie derivatives.}.   Neumann boundary conditions have been considered in recent works. See for instance a paper by Kofinas and Olea \cite{KO} on Lovelock anti-de Sitter gravity.   Our \sup $U^{\rs}_2$ does not apply to such spacetimes. 

\nnn d]   Jacobson and Myers used the first law of  thermodynamics to define    mass energy of Lovelock black holes. Mass-energy has, however, little to do with thermodynamics because     asymptotic spacetimes ignore the source of gravity.  A direct calculation of the mass-energy of Lovelock black-holes, independent of thermodynamic considerations, is  naturally provided by   \sup $U^{\rs}$. An   approach, similar to that of Jacobson and Myers, was used by Gibbons {\it et al} \cite{GLPP} to find the mass of Kerr-anti-de Sitter black holes. Needless to say the same mass-energy is obtained  with the KBL superpotential $U^{\rs}_1$, see \cite{DK}. 

\nnn d] Finally, we are still  wondering about the following problem. Noether's identity provides a conserved current   from a given Lagrangian. The conserved current is  the divergence of a non-unique superpotential. The form of that current (see   Section 1) tells us what physical meaning to attribute to closed surface integrals of the superpotential assuming   it has been properly chosen. With Silva's prescription, things work the other way round. One   calculates a unique \sup but one must look at its divergence to figure out the physical meaning of its total flux. Both methods are sound and well defined but what is the formal connexion?

 \vs
\Large{\bf Acknowledgements}
 \vskip .5 cm
 \normalsize 
 \setlength{\baselineskip}{20pt plus2pt}
  
 We thank Nathalie Deruelle for very useful discussions.

  \end{document}